\documentclass[twocolumn,usenatbib]{mn2e}
\usepackage{graphicx,natbib}
\usepackage{epsfig}

\newcommand{\beq}{\begin{equation}}
\newcommand{\eeq}{\end{equation}}
\def\lsim{\mathrel{\mathpalette\@versim<}}
\def\gsim{\mathrel{\mathpalette\@versim>}}

\def\ga{{\rm\thinspace gauss}}

\def\apj{\rm ApJ}

\title[Statistical classification of gamma-ray bursts]{Statistical classification of gamma-ray bursts based on the Amati relation}
\author[Y.-P. Qin and Z.-F. Chen]
{Yi-Ping Qin$^{1, 2, 3}$\thanks{E-mail:ypqin@126.com}, Zhi-Fu Chen$^{2, 1}$\\
$^{1}$ Center for Astrophysics, Guangzhou University, Guangzhou 510006, P. R. China\\
$^{2}$ Department of Physics and Telecommunication Engineering,
Baise University, Baise, Guangxi 533000, P. R. China\\
$^{3}$ Physics Department, Guangxi University, Nanning 530004, P. R.
China\\}
\begin{document}
\maketitle

\begin{abstract}
\noindent Gamma-ray bursts (GRBs) are believed to belong to two
classes and they are conventionally divided according to their
durations. This classification scheme is not satisfied due to the
fact that duration distributions of the two classes are heavily
overlapped. We collect a new sample (153 sources) of GRBs and
investigate the distribution of the logarithmic deviation of the
$E_p$ value from the Amati relation. The distribution possesses an
obvious bimodality structure and it can be accounted for by the
combination of two Gaussian curves. Based on this analysis, we
propose to statistically classify GRBs in the well-known $E_{p}$ vs.
$E_{iso}$ plane with the logarithmic deviation of the $E_p$ value.
This classification scheme divides GRBs into two groups: Amati type
bursts and non-Amati type bursts. While Amati type bursts well
follow the Amati relation, non-Amati type bursts do not. It shows
that most Amati type bursts are long duration bursts and the
majority of non-Amati type bursts are short duration bursts. In
addition, it reveals that Amati type bursts are generally more
energetic than non-Amati type bursts.
An advantage of the new classification is that the two kinds of
burst are well distinguishable and therefore their members can be
identified in certainty.
\end{abstract}


\section{Introduction}
Gamma-ray bursts (GRBs) are generally divided into two classes:
short and long-duration classes. The duration concerned is
always $T_{90}$ which is the time interval during which the burst
integrated counts increases from 5\% to 95\% of the total counts.
This classification scheme is based on the bimodality structure of
the $T_{90}$ distribution of the objects, where all the bursts are
likely to be separated at about 2 seconds (see, e.g., Kouveliotou et
al. 1993). When one replaces $T_{90}$ with $T_{50}$ (during
which the burst integrated counts increases from 25\% to 75\% of the
total counts), the bimodality structure also exists (see, e.g.,
Zhao et al. 2004). Generally, short duration bursts are
harder than long duration bursts. In the hardness ratio vs.
duration plane, short and long bursts were observed to distribute in
distinct domains ( Kouveliotou et al. 1993, Fishman \& Meegan 1995).
It was shown that the hardness ratio is correlated to the
duration for the whole GRB sample, but for each of the two classes
alone the two quantities are not correlated at all (see Qin et al.
2000). This statistical result strongly suggests that, while any
attempts to consider all GRBs as a single class might be
questionable, the existence of the two classes of GRBs is
convincing.

It is expected that different classes might have different
progenitors. Therefore, the classification of GRBs has always been
an essential task. Based on many years of investigation, most
researchers come to the consent that many shot bursts are produced
in the event of binary neutron star or neutron star-black hole
mergers, while many long bursts are caused by the massive star
collapsars (e.g., Eichler et al. 1989; MacFadyen \& Woosley 1999;
Paczynski 1986, 1998; Woosley 1993).

Thanks to the successful launch of the Swift satellite (Gehrels et
al. 2004), many advances in the research of GRBs have been achieved.
The most important achievement might be the fact that \textbf{a
large body of evidence favors} the two progenitor proposal for GRBs.
It has been continued to be reported that short bursts were found in
regions with lower star-formation rates, and no evidence of
supernovae to accompany them was detected (Barthelmy et al. 2005;
Berger et al. 2005; Hjorth et al. 2005). Meanwhile, long bursts were
found to be originated from star-forming regions in galaxies
(Fruchter et al. 2006), and in some of these events, supernovae were
detected to accompany the bursts (Hjorth et al. 2003; Stanek et al.
2003).

Short burst class and long burst class are conventionally divided by
$T_{90}$: those their $T_{90}$ being larger than 2 seconds
are classified as long bursts while the rest are classified as short
bursts. McBreen et al. (1994) showed that, the bimodal distribution
of GRBs can arise from two overlapped lognormal distributions. This
indicates that each of the two GRB populations might possess a
single lognormal duration distribution, and due to the overlap,
there would be a sufficient number of bursts that are mis-classified
by simply applying the criterion of $T_{90}=2s$.

The scenario that two overlapped lognormal distributions can account
for the duration distribution of the whole GRB sample was challenged
later by other investigations. Horvath (1998) found that, instead of
the two-Gaussian fit, the three-Gaussian fit is more likely to be
able to account for the duration distribution of all BATSE bursts.
Although if there exists a third class of GRBs is stills a subject
of debate, the evidence that the two-Gaussian fit alone cannot
account for the duration distribution of all GRBs is obvious.
In fact, $T_{90}$ is not an intrinsic property of a burst or
a population of bursts. For a more robust investigation, one should
rely on the cosmological rest-frame duration (see the definition of
$T_{90,r}$ below), where the effect of cosmological redshift has
been corrected. Unfortunately, the redshift information is not
available for most BATSE bursts, and hence in the corresponding
analysis this effect can not be taken into account. However, in our
analysis below, redshifts of the bursts are known, and therefore we
will use $T_{90,r}$ instead of $T_{90}$.

In fact, voices questioning the duration classification scheme have
become stronger in recent years. Gehrels et al. (2006) reported that
the duration of GRB 060614 is long but its behavior is similar to
short duration bursts (for example, very deep optical observations
of this source exclude an accompanying supernova). Based on this
fact, they even asked if there exists a new GRB classification
scheme that straddles both long and short duration bursts. Similar
observational results were reported by different groups in nearly
the same time (see, e.g., Gal-Yam et al. 2006; Fynbo et al. 2006;
Della Valle et al. 2006). For some short duration bursts, soft
extended emission and late X-ray flares were observed, indicating
that these sources might not really short (see, e.g., Barthelmy et
al. 2005; Norris \& Bonnell 2006).

In the recent few years, some attempts of revealing new statistical
properties associated with the classification of bursts as well as
introducing new classification schemes have continued to be made.
Zhang et al. (2007) proposed that GRBs should be classified into
Type I (typically short and associated with old populations) and
Type II (typically long and associated with young populations)
groups. This type of classification is charming, but the goal of
dividing individual bursts into the distinct groups is hard to
realize. Lu et al. (2010) introduced a new parameter to classify
GRBs. In their efforts, they regarded those long GRBs with
``extended emission'' being short ones if the bursts are really
``short'' without the ``extended emission''. In this way, they found
a clear bimodal distribution of the parameter. Goldstein et al.
(2010) found the distribution of the ratio of Epeak/Fluence bearing
a bimodality structure in the complete BATSE 5B spectral catalog,
which corresponds directly to the conventional short and long burst
classes. However, the overlap of the distributions of the two groups
of bursts is seen to be as heavy as that shown in the duration
distributions. Qin et al. (2010) proposed to modify the conventional
duration classification scheme by separating the conventional short
and long duration bursts in different softness (or hardness) groups.
While this method seems reasonable, the improvement would not be
applicable if the size of samples is not large enough.

Just as was mentioned above, one can verify that, two
distinct smooth curves (e.g., two Gaussian curves) accounting for
the duration distributions of the short and long burst classes are
sufficiently overlapped. This makes the duration classification
scheme an unsatisfied one. Unfortunately, the overlap of other
parameters (e.g., the hardness ratio or the peak energy) is much
heavier than that of the duration. Although it is much beyond being
satisfactory, the duration classification scheme is still the most
popular one up to day. Thus, It is desirable that a better
alternative of the classification can be established in the near
future.

Based on a sample of BeppoSAX GRBs with known redshift, Amati et al.
(2002) discovered a tight relation between the cosmological
rest-frame spectrum peak energy and the isotropic equivalent
radiated energy, which is now known as the Amati relation. This soon
triggered a series of relevant researches (e.g., Amati 2006, 2010;
Amati et al. 2007, 2008, 2009; Piranomonte et al. 2008; Ghirlanda et
al. 2008, 2009; Gruber et al. 2011; Zhang et al. 2012).

There have been debates about the existence of the Amati
relation as an intrinsic property of GRBs. Some authors pointed out
that the relation might arise from observational selection effects
(e.g., Band \& Preece 2005; Butler et al. 2007, 2009;  Nakar \&
Piran 2005). Other authors argued that, to form the relation,
selection effects could only play a marginal role (see, e.g., Amati
et al. 2009; Bosnjak et al. 2008; Ghirlanda et al. 2005, 2008; Krimm
et al. 2009; Nava et al. 2008). Recently, Butler et al. (2010)
derived a GRB world model from their data, and based on the model
they reproduced the observables from both Swift and pre-Swift
satellites. In their analysis, a real, intrinsic correlation between
the two quantities is confirmed, but they pointed out that the
correlation is not a narrow log - log relation and its observed
appearance is strongly detector-dependent. Our data (see the
analysis below) show that the Amati relation is real, although it
might be, at least in part, introduced by observational bias.

In a subsequent analysis with a much larger sample, Amati (2006)
reported that subenergetic GRBs (such as GRB 980425) and short GRBs
are found to be inconsistent with the correlation between the
cosmological rest-frame spectrum peak energy and the isotropic
equivalent radiated energy, indicating that this phenomenon might be
a powerful tool for discriminating different classes of GRBs and
understanding their nature and differences. Recently, Zhang et al.
(2012) selected some short bursts and disregarded those subenergetic
GRBs concerned by Amati (2006). They reported that, for these short
bursts alone, there does exist a tight relation between the two
quantities, which is quite different from the conventional Amati
relation.

As the Amati relation is real and the number of GRBs with known
redshift has become larger and larger in recent years, it might be
possible now to use the relation to distinguish members of distinct
GRB classes statistically. This motivates our analysis below.

In Section 2, the collection of GRBs with known redshift is
presented and a statistical analysis is performed to check if there
exits an appropriate quantity to separate the bursts into different
groups. Based on this analysis, we present a new classification
scheme in Section 3. In Section 4, we compare the new classification
scheme and the conventional scheme. A summary and discussion are
presented in Section 5.

Throughout this paper, we adopt the following cosmological
parameters: $H_{0}=70kms^{-1}Mpc$, $\Omega_{M}=0.3$ and
$\Omega_{\Lambda}=0.7$.

\section{Data and analysis}
We only consider GRBs with measured redshift, up to May
2012, including sources observed by various instruments. In
addition, some other quantities are required. In fact, included in
our sample are merely those GRBs with the following quantities
available: redshift $z$, spectrum peak energy $E_p$, isotropic
equivalent radiated energy $E_{iso}$, and duration $T_{90}$. We get
153 bursts in total. Sources of our sample are listed in Table 1.

Let $E_{p,r}\equiv(1+z)E_{p}$ being the cosmological
rest-frame $\nu f_{\nu}$ spectrum peak energy (in brief, the
rest-frame peak energy), in units of $kcV$, and $E_{iso}$ being the
isotropic equivalent radiated energy (in brief, the isotropic
energy), in units of $10^{52}erg$. In the following, when $E_{p,r}$
and $E_{iso}$ are used in any analysis, they stand for their
observational values (see, e.g., Table 1). The Amati relation can be
expressed as follows (see Amati 2006):
\begin{equation}
E_{p,r,pre}=K\times E^{m}_{iso},
\end{equation}
where subscript $pre$ means ``predicted'', $K$ and $m$ are constants
obtained by fits. In order to avoid notation confusion in the
analysis below, we use $E_{p,r,pre}$ to describe the predicted value
of the rest-frame peak energy, determined by the Amati relation when
$E_{iso}$ is provided.

To check if a burst obey or betray the Amati relation, we follow
Amati (2006) to consider the logarithmic deviation of the $E_p$
value from the Amati relation that serves as a datum line in the
$E_{p}$ vs. $E_{iso}$ plane. The Amati relation adopted as
the datum line in this paper is that obtained by Amati et al.
(2008): $E_{p,r,pre}=94 \times E^{0.57}_{iso}$. Thus, the
logarithmic deviation of the $E_p$ value considered in this paper is
$log E_{p,r} - log 94 - 0.57 log E_{iso}$, where $K=94$ is different
from that adopted in Figure 4 of Amati (2006). (Note that, the Amati
relation is improved in Amati et al. 2008 with a much larger sample
compared with that in Amati 2006).

\begin{figure}
\center
\includegraphics[width=7.5cm,height=6.5cm]{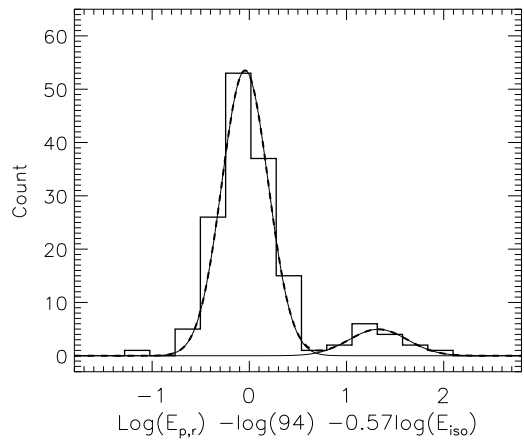}\\
\caption{Distribution of the logarithmic deviation of the $E_p$
value of our sample (153 sources), where the deviation is calculated
by $log E_{p,r} - log 94 - 0.57 log E_{iso}$. The thick dash line
represents a two-Gaussian fit, and the two thin solid lines (heavily
overlapped by the thick dash line) stand for the two Gaussian curves
respectively. There is a dip at about 0.7. The number of the sources
located at the left hand side of the dip is 137, while that of the
rest is 16.}
\end{figure}

Displayed in Fig. 1 is the distribution of the logarithmic deviation
of the $E_p$ value of our sample. Unlike that shown in Figure 4 of
Amati (2006) (where only long GRBs and X-ray flashes are
considered), the distribution in our sample shows an obvious
bimodality structure. It reveals that there are two Gaussian
distributions that form the bimodality structure, and the overlap of
the two distributions is not heavy (see Fig. 7 for a comparison with
the $T_{90}$ distribution of the same sample).

We perform a two-Gaussian fit to the distribution of the logarithmic
deviation of the $E_p$ value (i.e., the distribution of $log E_{p,r}
- log 94 - 0.57 log E_{iso}$), and obtain $\rm \sigma=0.239$ and a
central value of -0.044 for the first Gaussian curve, and $\rm
\sigma=0.300$ and a central value of 1.327 for the second Gaussian
curve, with the reduced $\rm \chi^2$ of the fit being $\rm
\chi^2_{dof}=16.649$.

We find that, 100 percent (16/16) of the bursts accounted for by the
second curve are located beyond the $3\sigma$ range of the first
curve, and 91.2 percent (125/137) of the bursts accounted for by the
first curve are located beyond the $3\sigma$ range of the second
curve, which indicates that the overlap of the two Gaussian
distributions is very light.

\section{New classification}
The bimodality structure shown in Fig. 1 favors the assumption that
there are two distinct classes of GRBs. If we believe that each of
the two Gaussian distributions obtained above corresponds to one of
the two classes, then the figure indicates that while members of one
class clustering around the Amati relation (represented by the zero
value of the deviation; see Fig. 1), sources of the other class are
located far away from the relation. This encourages us to use a
logarithmic deviation of the $E_p$ value to set apart the classes of
GRBs.

\begin{figure}
\center
\includegraphics[width=7.5cm,height=6.5cm]{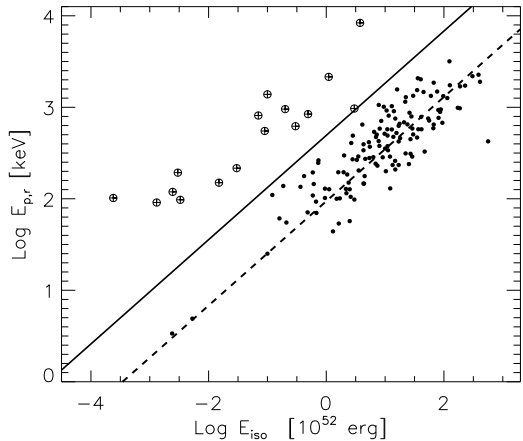}\\
\caption{Classification of the 153 GRBs in the $E_{p}$ vs. $E_{iso}$
plane. The dash line represents the Amati relation, and the solid
line represents the criterion curve of the new classification,
equation (2). Filled circles stand for Amati type bursts (137
sources), and open circles plus pluses reprensent non-Amati type
bursts (16 sources).}
\end{figure}

According to Fig. 1 and the fitting curve, we assign the logarithmic
deviation of the $E_p$ value located at the dip between the two
peaks of the fitting curve as the criterion to classify members of
the two groups. The dip is at 0.72. It corresponds to the following
curve in the $E_{p}$ vs. $E_{iso}$ plane:
\begin{equation}
E_{p,r,pre}=10^{0.72} \times 94 \times E^{0.57}_{iso}.
\end{equation}
Sources located under this curve in the $E_{p}$ vs. $E_{iso}$ plane
are classified as Amati type bursts, and that located above this
curve are classified as non-Amati type bursts. Or, in terms of the
logarithmic deviation of the $E_p$ value, GRBs with $log E_{p,r} -
log 94 - 0.57 log E_{iso} < 0.72$ are classified as Amati type
bursts, otherwise they are classified as non-Amati type bursts.
Shown in Fig. 2 is the result of the classification.

\begin{figure}
\center
\includegraphics[width=7.5cm,height=6.5cm]{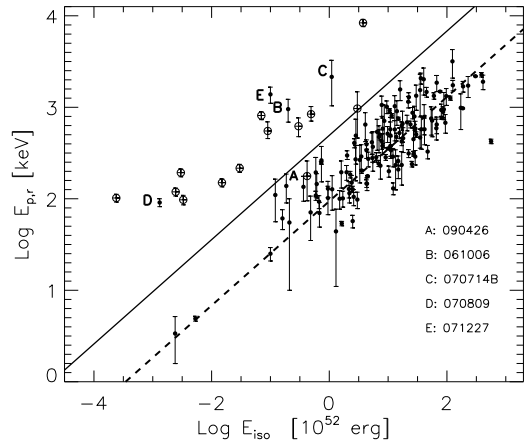}\\
\caption{Distributions of the short (open circles plus pluses) and
long (filled circles) bursts of the 153 GRBs in the $E_{p}$ vs.
$E_{iso}$ plane. The dash line represents the Amati relation, and
the solid line represents the criterion curve of the new
classification, equation (2). Only one short burst (GRB 090426) is
located below the criterion curve, and four long bursts (GRB
061006,GRB 070714B, GRB 071227 and GRB 070809) are located above the
curve.}
\end{figure}

To check how different duration (short or long) bursts are
influenced by this classification, let us follow the conventional
method to classify the bursts by duration. Since the redshifts of
bursts are available, we modify the conventional duration
classification criterion by replacing $T_{90}=2s$ with
$T_{90,r}=1s$, where $T_{90,r}\equiv T_{90}/(1+z)$ is the
cosmological rest-frame duration (shortly, rest-frame duration). We
divide GRBs into two groups by assigning bursts with $T_{90,r}>1$
for one group (long bursts) and bursts with $T_{90,r}\leq1$ for the
other group (short bursts). In this way we get 13 short bursts in
total. The reason for taking $T_{90,r}=1s$ as the duration criterion
is that, it corresponds to $T_{90}=2s$ when $z=1$. In our sample,
when bursts are divided by $T_{90}=2s$ we get 16 short ones.
Therefore, to get a sample of short bursts, the criterion of
$T_{90,r}=1s$ is more conservative than the conventional one, that
of $T_{90}=2s$.

Distributions of these two groups of bursts (short and long bursts)
are displayed in Fig. 3. We find that short bursts are mainly
(12/13, 92.3\%) located in the non-Amati type burst region while
long bursts are preferentially (136/140, 97.1\%) distributed in the
Amati type burst domain.

\begin{figure}
\center
\includegraphics[width=7.5cm,height=6.5cm]{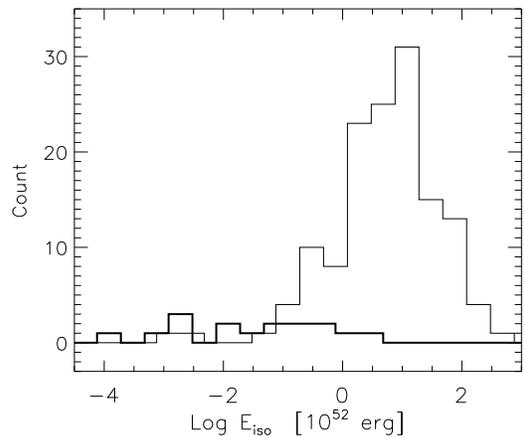}\\
\caption{Distributions of $\rm E_{iso}$ for the two newly classified
groups of bursts. The thick solid line stands for the group of
non-Amati type bursts, and the thin solid line corresponds to the
group of Amati type bursts.}
\end{figure}

\begin{figure}
\center
\includegraphics[width=7.5cm,height=6.5cm]{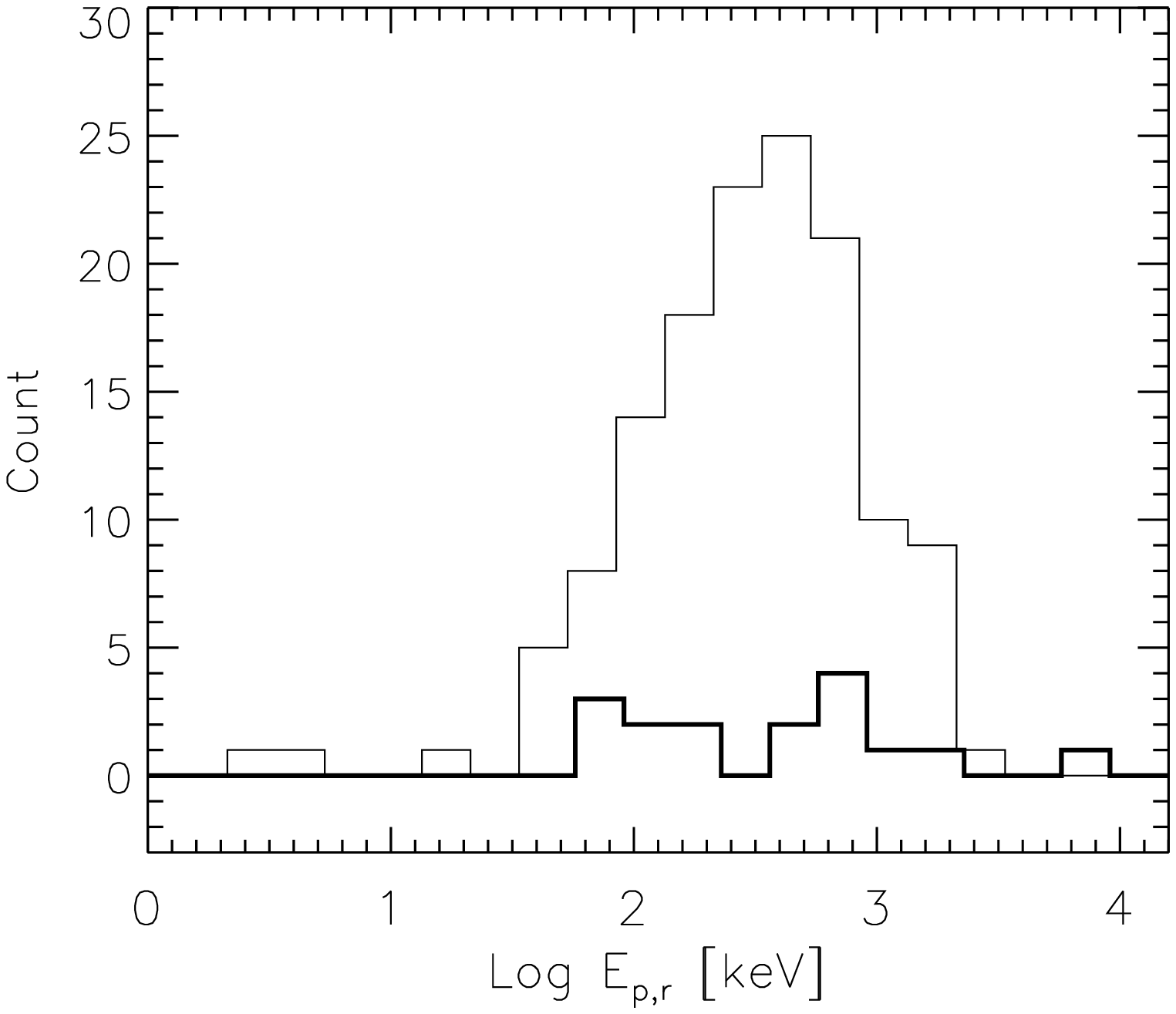}\\
\caption{Distributions of $\rm E_{p,r}$ for the two newly classified
groups of bursts. The meanings of lines are the same as that in Fig.
4.}
\end{figure}

\begin{figure}
\center
\includegraphics[width=7.5cm,height=6.5cm]{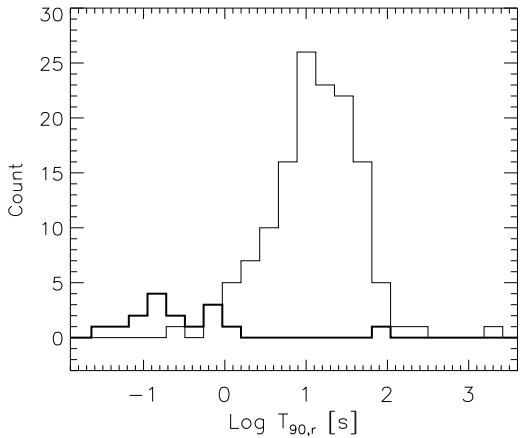}\\
\caption{Distributions of $\rm T_{90,r}$ for the two newly
classified groups of bursts. The meanings of lines are the same as
that in Fig. 4.}
\end{figure}

Shown in Figs. 4, 5, and 6 are the distributions of $E_{iso}$,
$E_{p,r}$, and $T_{90,r}$ respectively for the two newly classified
groups of bursts. We find from these figures that Amati type bursts
are generally longer and more energetic. Unlike in the case of the
conventional duration classification scheme, the two newly
classified groups of bursts do not show significant difference in
the distribution of $E_{p,r}$. One can also observe this in Fig. 2.
It has been known for a long time that sources of the conventional
short burst class are generally harder than those of the
conventional long burst class. At least with the current sample (153
sources), this difference is relatively mild if sources are divided
with the new classification scheme.

\section{Comparison}
Let us compare the new classification scheme, the scheme based on
the logarithmic deviation of the $E_p$ value (shortly, the peak
energy deviation classification scheme), with the conventional
duration classification scheme.

\begin{figure}
\center
\includegraphics[width=7.5cm,height=6.5cm]{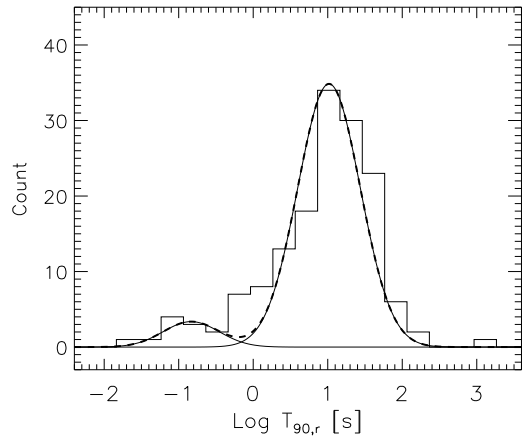}\\
\caption{Distribution of $\rm T_{90,r}$, of our sample. The meanings
of lines are the same as that in Fig. 1.}
\end{figure}

Shown in Fig. 7 is the distribution of the rest-frame duration, of
our sample. The well-known bimodality structure is observed. As is
already known, it is unlikely that the bimodality structure
distribution arises from the combination of two Gaussian
distributions.

As done in the case of the peak energy deviation classification
scheme, we fit the duration distribution of the sample with the
combination of two Gaussian functions and get $\rm \sigma=0.375$ and
a central value of -0.833 for the first Gaussian curve, and $\rm
\sigma=0.419$ and a central value of 1.017 for the second Gaussian
curve, with the reduced $\rm \chi^2$ of the fit being $\rm
\chi^2_{dof}=29.826$. We find that the resulting reduced $\rm
\chi^2$ value (29.826) is much larger than that (16.649) of the new
classification scheme. As a key parameter of classification to
separate two groups of sources, one always expects its distribution
to possess a bimodality structure that arises from the combination
of two perfect Gaussian curves. In this aspect, the logarithmic
deviation of the $E_p$ value acts much better than the duration
does.

In addition, we perform a linear fit to the $E_{p,r}$ and $E_{iso}$
data of the two kinds of duration bursts, those with $T_{90,r}>1s$
and that with $T_{90,r}\leq1s$, in our sample. It yields:
\begin{equation}
Log E_{p,r}=(2.12\pm0.15) + (0.46\pm0.11)Log E_{iso}
\end{equation}
for bursts with $T_{90,r}>1s$ ($\rm N=140,~r=0.754,~P<10^{-27}$),
and
\begin{equation}
Log E_{p,r}=(2.95\pm0.69) + (0.32\pm0.36)Log E_{iso}
\end{equation}
for bursts with $T_{90,r}\leq1s$ ($\rm N=13,~r=0.846,~P=0.0003$).

As a comparison, the same analysis is performed in the case of the
new classification scheme. That produces:
\begin{equation}
Log E_{p,r}=(2.06\pm0.16) + (0.51\pm0.12)Log E_{iso}
\end{equation}
for Amati type bursts ($\rm N=137,~r=0.831,~P<10^{-36}$), and
\begin{equation}
Log E_{p,r}=(3.16\pm0.65) + (0.39\pm0.33)Log E_{iso}
\end{equation}
for non-Amati type bursts ($\rm N=16,~r=0.912,~P<10^{-7}$).

\begin{figure}
\center
\includegraphics[width=7.5cm,height=6.5cm]{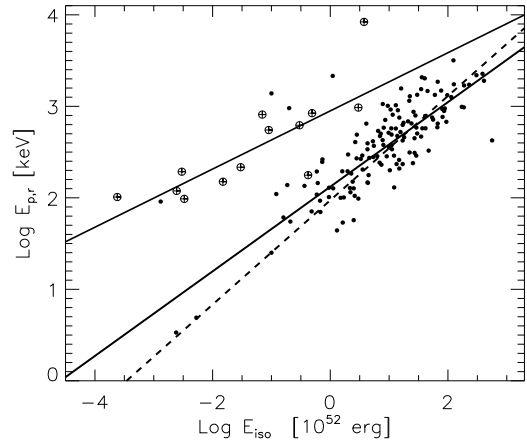}\\
\caption{Results of correlation analysis between $E_{p,r}$ and
$E_{iso}$ for short and long bursts of our sample, divided by
$T_{90,r}=1s$. The upper solid line represents the linear fit to the
short bursts, and the lower solid line represents the linear fit to
the long bursts. See Fig. 3 for the meanings of the dash line and
other symbols.}
\end{figure}

\begin{figure}
\center
\includegraphics[width=7.5cm,height=6.5cm]{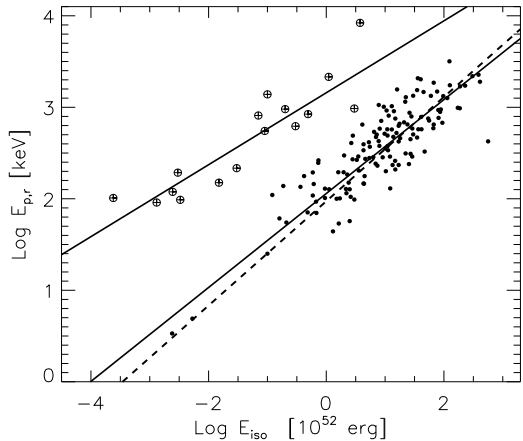}\\
\caption{Results of correlation analysis between $E_{p,r}$ and
$E_{iso}$ for the two groups of bursts of our sample, divided by the
newly classification scheme. The upper solid line represents the
linear fit to the non-Amati type bursts, and the lower solid line
represents the linear fit to the Amati type bursts. See Fig. 2 for
the meanings of the dash line and other symbols.}
\end{figure}

The results are displayed in Figs. 8 and 9 respectively. It suggests
that, if we believe that bursts of the same class should follow the
same relationship between $E_{p,r}$ and $E_{iso}$, as hinted by the
discovery of Amati et al. (2002), then the duration of bursts cannot
appropriately separate the two classes. In this aspect and in terms
of statistics, the logarithmic deviation of the $E_p$ value is much
more preferential than the duration.

\section{Summary and discussion}
We collected GRBs with measured redshift, spectrum peak
energy, isotropic equivalent radiated energy, and duration from
literature up to May 2012, including sources observed by various
instruments, and get 153 GRBs in total. With this sample, we
investigate the distribution of the logarithmic deviation of the
$E_p$ value from the Amati relation. The distribution exhibits an
obvious bimodality structure. A fit to the data shows that the
distribution of the deviation can be accounted for by the
combination of two Gaussian curves, and the two curves are well
separated. Based on this, we propose to statistically classify GRBs
in the $E_{p}$ vs. $E_{iso}$ plane with the logarithmic deviation of
the $E_p$ value. According to this classification scheme, bursts are
divided into two groups: Amati type bursts and non-Amati type
bursts. A statistical interpretation of this classification is that,
Amati type bursts well follow the Amati relation, but non-Amati type
bursts do not. Our analysis reveals that Amati type bursts are
generally longer and more energetic and non-Amati type bursts are
generally shorter and less energetic. After comparing the new
classification scheme with the conventional scheme we find that, in
terms of statistics, the logarithmic deviation of the $E_p$ value
acts much better in the classification routine than the duration
does. Since the overlap of the distributions of the logarithmic
deviation of the $E_p$ value is light, the two groups of bursts so
divided are well distinguishable and therefore their members can be
identified in certainty.

From Fig. 7, one might observe that, taking $T_{90,r}=1s$ as the
duration criterion might not be so appropriate since the dip between
the two peaks of the bimodality structure is located at the position
of a much smaller duration value. How the short and long bursts act
if we divide them at this position?

\begin{figure}
\center
\includegraphics[width=7.5cm,height=6.5cm]{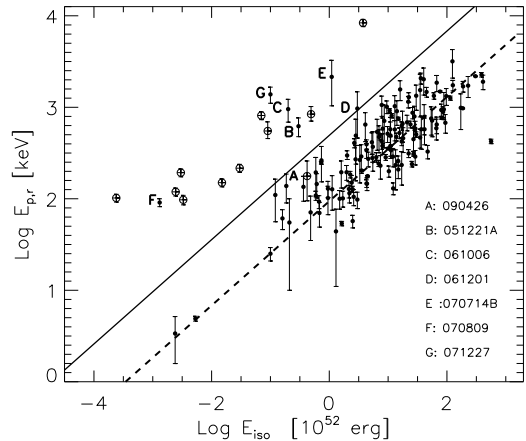}\\
\caption{Distributions of the short and long bursts of the 153 GRBs
in the $E_{p}$ vs. $E_{iso}$ plane, classified by the duration
criterion of $T_{90,r}=0.63s$. The meanings of the lines and symbols
are the same as that in Fig. 3. There is only one short bursts (GRB
090426) located below the criterion curve, while there are six long
bursts (GRB 051221A, GRB 061006, GRB 061201, GRB 070714B, GRB
070809, GRB 070809 and GRB 071227) located above the curve.}
\end{figure}

\begin{figure}
\center
\includegraphics[width=7.5cm,height=6.5cm]{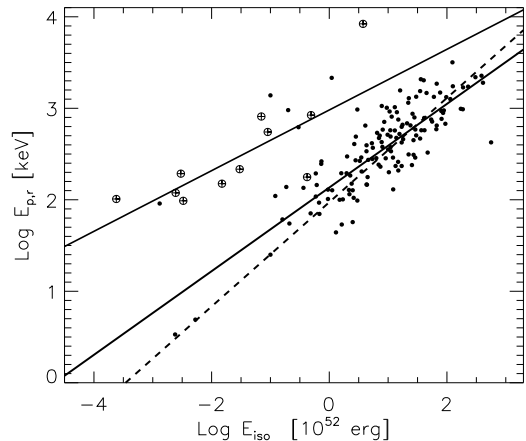}\\
\caption{Results of correlation analysis between $E_{p,r}$ and
$E_{iso}$ for short and long bursts of our sample, classified by the
duration criterion of $T_{90,r}=0.63s$. See Fig. 8 for the meanings
of the lines and symbols.}
\end{figure}

According to Fig. 7, this position is at $T_{90,r}=0.63s$. Let us
divide bursts into two groups by taking $T_{90,r}=0.63s$ as the
duration criterion. In this way, we get 11 short bursts and 142 long
bursts. The distributions of these groups of bursts in the $E_{p}$
vs. $E_{iso}$ plane are shown in Fig. 10. We find that 90.9\%
(10/11) of this kind of short burst are classified as non-Amati type
bursts, and 95.8\% (136/142) of this kind of long burst are
classified as Amati type bursts. We repeat the above linear analysis
for these two groups and get
\begin{equation}
Log E_{p,r}=(2.13\pm0.15) + (0.46\pm0.11)Log E_{iso}
\end{equation}
for bursts with $T_{90,r}>0.45s$ ($\rm N=142,~r=0.731,~P<10^{-25}$),
and
\begin{equation}
Log E_{p,r}=(2.98\pm0.85) + (0.33\pm0.41)Log E_{iso}
\end{equation}
for bursts with $T_{90,r}\leq0.45s$ ($\rm N=11,~r=0.809,~P=0.003$).
The correlation analysis results are shown in Fig. 11. It suggests
that the duration criterion changing from $T_{90,r}=1s$ to
$T_{90,r}=0.63s$ does not give rise to a significantly different
result.

We notice that our linear analysis result for short bursts is quite
different from that obtained by Zhang et al. (2012). This must be
due to the fact that our short burst sample (even in the case of
adopting the duration criterion of $T_{90,r}=0.63s$) is larger than
that of Zhang et al. (2012). Note that, short burst GRB 090426
($T_{90}=1.2s$, $T_{90,r}=0.33s$) was omitted in Zhang et al.
(2012), and this burst is just located within the Amati type burst
domain (see Fig. 10).

Although certain answers might not be available currently, we still
like to raise some questions associated with the new classification
scheme, in order to urge more relevant investigations. a) Beside the
statistical interpretation mentioned above, do there exist any
mechanisms accounted for the two classes? What mechanisms or
physical conditions the Amati relation reveals? b) How about the two
newly classified groups of bursts are related to the Type 1 and Type
II bursts? c) What is the role this new classification scheme plays?
How it relates to the conventional duration classification scheme?
Can both schemes be combined to find intrinsically different groups?
Or, other classification schemes should be involved?

Amati et al. (2009) pointed out that the Amati relation can be
explained by the non-thermal synchrotron radiation scenario, e.g.,
by assuming that the minimum Lorentz factor, and the normalization
of the power-law distribution of the radiating electrons do not vary
significantly from burst to burst or when imposing limits on the
slope of the correlation between the fireball bulk Lorentz factor,
and the burst luminosity (Lloyd et al. 2000; Zhang \& Meszaros
2002). Is the relation short bursts follow due to the same
mechanism? If so, why are the two relations different?

As discussed in Amati (2010), those long bursts to be seen off-axis
could betray the conventional Amaiti relation and become outliers.
The fact that short GRBs do not follow the Amati relation might be
due to their different progenitors (likely mergers) or the
difference of the circum-burst environment and the main emission
mechanisms. Why are these outliers and short bursts located
in the same region in the $E_{p}$ vs. $E_{iso}$ plane and following
the same relation? Perhaps they share some common physical
conditions that are different from what most long bursts possess.

GRB 060614 is a typical burst which lasts long enough but it is not
obviously associated with SN. As this burst is found to be
consistent with the Amati relation as most GRB/SN events, Amati et
al. (2007) suggested that the position in the $E_{p}$ vs. $E_{iso}$
plane of long GRBs does not critically depend on the progenitor
properties. However, when taking into account only its first spike,
GRB 060614 will shift from the Amati burst domain to non-Amati burst
domain (see, e.g., Amati 2010). If one believes that GRB 060614 is a
Type I burst, then one must come to this conclusion: at least in
general cases, Type I and Type II bursts are not necessarily to be
well separated in the $E_{p}$ vs. $E_{iso}$ plane. Or, Amati bursts
are not necessary to be Type II sources and non-Amati bursts are not
necessary to be Type I GRBs. Perhaps when special treatment such as
considering only the first spike of bursts is employed, the
conclusion will be changed.

If it is true that Type I and Type II bursts are not necessarily to
be well separated in the $E_{p}$ vs. $E_{iso}$ plane, then the peak
energy deviation classification scheme alone would not be able to
classify bursts with different progenitors. In this case, other
classification schemes should be involved. Perhaps one can combine
several schemes to set apart these bursts. If so, combination of
both the peak energy deviation classification scheme and the
conventional duration classification scheme might give rise to a
much better result.

\begin{figure}
\center
\includegraphics[width=7.5cm,height=6.5cm]{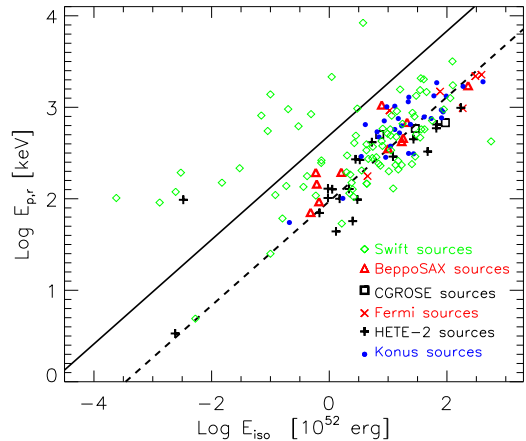}\\
\caption{Distributions of the bursts detected by various instruments
in the $E_{p,r}$ vs. $E_{iso}$ plane.}
\end{figure}

As mentioned above, the Amati relation might probably be
affected by observational bias. Illustrated in Fig. 12 are the
distributions of the bursts detected by various instruments in the
$E_{p,r}$ vs. $E_{iso}$ plane. We find that the domains of the
distributions of the bursts observed by different instruments are
not fully coincident. Especially, difference between the domain of
the bursts observed by Swift and that of the bursts observed by
other instruments is quite obvious. There does exist instrument
bias. A robust analysis of statistical classification requires
samples without any observational bias, which seems not being
available currently.

\begin{figure}
\center
\includegraphics[width=7.5cm,height=6.5cm]{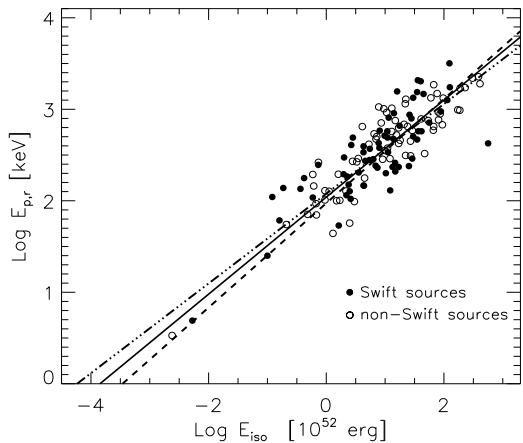}\\
\caption{Results of correlation analysis between $E_{p,r}$ and
$E_{iso}$ for the Amati type bursts detected by Swift (Swift
sources) and other instruments (non-Swift sources) respectively. The
dash line represents the Amati relation, the dot dot dash line
represents the linear fit to the Swift sources, and the solid line
stands for the linear fit to the non-Swift sources.}
\end{figure}

From Fig. 12 we find that the bias introduced by the
observation of Swift comes mainly from the joining of most of the
non-Amati bursts (including the majority of conventional short
bursts and the outliers of conventional long bursts; see Fig. 3).
According to the above analysis, we regard this as a contribution of
Swift to the new classification scheme. This is favored by the
following fact: when one considers only the Amati type bursts (those
under the solid line in Fig. 2), one would find that the bias of
Swift is mild. We perform correlation analysis between $E_{p,r}$ and
$E_{iso}$ for the Amati type bursts detected by Swift and other
instruments respectively. The analysis produces:
\begin{equation}
Log E_{p,r}=(2.08\pm0.22) + (0.49\pm0.18)Log E_{iso}
\end{equation}
for the Amati type Swift bursts ($\rm
N=67,~r=0.798,~P<10^{-16}$), and
\begin{equation}
Log E_{p,r}=(2.04\pm0.22) + (0.53\pm0.16)Log E_{iso}
\end{equation}
for the Amati type non-Swift bursts ($\rm
N=70,~r=0.847,~P<10^{-20}$). Presented in Fig. 13 are the results of
the analysis. It shows that, for the Amati type bursts alone, no
significant observational bias of Swift is observed.

In fact, for a complete analysis, one cannot rely on the
bursts observed only by a single instrument to discuss the
classification scheme. Instead, one should rely on all the bursts
that are observed by various instruments over the same area of sky
and during the same interval of time. This might be a great task
performed later. At present, to investigate the statistical
classification, we prefer all available bursts rather than only
those observed by a single instrument, since any instruments might
introduce (strong or mild) bias. Currently, no one exactly knows how
a complete sample would affect the statistical analysis above. Based
on Fig. 12, we suspect that, when the number of bursts observed by
all instruments increases, the clustering around the Amati relation
might become stronger and this will give rise to a well definition
of the Amati type bursts. In return, this will also be helpful to
distinguish the non-Amati type bursts.

An important difference between the original duration
classification and the one presented here is that the original was
conceived as a discriminator in the observer frame. The observed
duration is measured in the observer frame and is influenced by the
cosmological redshift. Therefore, to investigate intrinsic
properties of the sources, one needs to remove this effect from the
quantities concerned so that one can deal with them in the source
frame. This is the reason why we use $E_{p,r}$ and $T_{90,r}$ to
replace $E_{p}$ and $T_{90}$ respectively. In addition, to calculate
$E_{iso}$, one needs to know redshift as well. Obviously, the
information of redshift is essential for a deep investigation of
GRBs. We expect more and more bursts with known redshift being well
observed in the near future.

\vspace{6mm} We thank the anonymous referee for his/her helpful
suggestions that improve this paper greatly. This work was supported
by the National Natural Science Foundation
of China (No. 11073007) and the Guangzhou technological project (No. 11C62010685).\\

\setcounter{table}{0}
\begin{table*}
\caption{Parameters of prompt emission of GRBs with measured
redshifts.}
\begin{tabular}{llllllllllll}
\hline
GRB&z&$T_{90}$&$E_{iso}$&$E_{p,r}$&Instruments&Refs&Note\\
&&&$10^{52}$ erg&keV&&&\\
\hline
970228  &   0.695   &   80  &   1.6 $\pm$   0.12        &   195 $\pm$                   64      &   BeppoSAX    &   15  &   A   \\
970508  &   0.835   &   20  &   0.612   $\pm$   0.13        &   145 $\pm$                   43      &   BeppoSAX    &   15  &   A   \\
970828  &   0.958   &   146.59  &   29  $\pm$   3       &   586 $\pm$                   117     &   CGROSE  &   15  &   A   \\
971214  &   3.42    &   35  &   21  $\pm$   3       &   685 $\pm$                   133     &   BeppoSAX    &   15  &   A   \\
980326  &   1   &   9   &   0.482   $\pm$   0.09        &   71  $\pm$                   36      &   BeppoSAX    &   15  &   A   \\
980613  &   1.096   &   20  &   0.59    $\pm$   0.09        &   194 $\pm$                   89      &   BeppoSAX    &   15  &   A   \\
980703  &   0.966   &   102.37  &   7.2 $\pm$   0.7     &   503 $\pm$                   64      &   CGROSE  &   15  &   A   \\
990123  &   1.6 &   100 &   229 $\pm$   37      &   1724    $\pm$                   446     &   BeppoSAX    &   15  &   A   \\
990506  &   1.3 &   220.38  &   94  $\pm$   9       &   677 $\pm$                   156     &   CGROSE  &   15  &   A   \\
990510  &   1.619   &   75  &   17  $\pm$   3       &   423 $\pm$                   42      &   BeppoSAX    &   15  &   A   \\
990705  &   0.843   &   42  &   18  $\pm$   3       &   459 $\pm$                   139     &   BeppoSAX    &   15  &   A   \\
990712  &   0.434   &   20  &   0.67    $\pm$   0.13        &   93  $\pm$                   15      &   BeppoSAX    &   15  &   A   \\
991208  &   0.706   &   60  &   22.3    $\pm$   1.8     &   313 $\pm$                   31      &   Konus   &   15  &   A   \\
991216  &   1.02    &   24.9    &   67  $\pm$   7       &   648 $\pm$                   134     &   GRO/KW  &   15  &   A   \\
000131  &   4.5 &   110.1   &   172 $\pm$   30      &   987 $\pm$                   416     &   GRO/KW  &   15  &   A   \\
000210  &   0.846   &   20  &   14.9    $\pm$   1.6     &   753 $\pm$                   26      &   Konus   &   15  &   A   \\
000418  &   1.12    &   30  &   9.1 $\pm$   1.7     &   284 $\pm$                   21      &   Konus   &   15  &   A   \\
000911  &   1.06    &   500 &   67  $\pm$   14      &   1856    $\pm$                   371     &   Konus   &   15  &   A   \\
000926  &   2.07    &   25  &   27.1    $\pm$   5.9     &   310 $\pm$                   20      &   Konus   &   15  &   A   \\
010222  &   1.48    &   130 &   81  $\pm$   9       &   766 $\pm$                   30      &   Konus   &   15  &   A   \\
010921  &   0.45    &   24.6    &   0.95    $\pm$   0.1     &   129 $\pm$                   26      &   HETE每2 &   15  &   A   \\
011121  &   0.36    &   75  &   7.8 $\pm$   2.1     &   1060    $\pm$                   265     &   BeppoSAX    &   9   &   A   \\
020124  &   3.198   &   78.6    &   27  $\pm$   3       &   448 $\pm$                   148     &   HETE每2 &   15  &   A   \\
020127  &   1.9 &   17.6    &   3.5 $\pm$   0.1     &   290 $\pm$                   100     &   HETE每2 &   19  &   A   \\
020405  &   0.695   &   60  &   10  $\pm$   0.9     &   354 $\pm$                   10      &   BeppoSAX    &   15  &   A   \\
020813  &   1.25    &   90  &   66  $\pm$   16      &   590 $\pm$                   151     &   HETE每2 &   15  &   A   \\
020819B &   0.41    &   46.9    &   0.68    $\pm$   0.017       &   70  $\pm$                   21      &   HETE每2 &   15  &   A   \\
020903  &   0.25    &   10  &   0.0024  $\pm$   0.0006      &   3.37    $\pm$                   1.79        &   HETE每2 &   15  &   A   \\
021004  &   2.3 &   77.1    &   3.3 $\pm$   0.4     &   266 $\pm$                   117     &   HETE每2 &   15  &   A   \\
021211  &   1.01    &   2.41    &   1.12    $\pm$   0.3     &   127 $\pm$                   52      &   HETE每2 &   15  &   A   \\
030226  &   1.98    &   76.8    &   12.1    $\pm$   1.3     &   289 $\pm$                   66      &   HETE每2 &   15  &   A   \\
030323  &   3.37    &   32.6    &   2.8 $\pm$   0.9     &   270 $\pm$                   113     &   HETE每2 &   10  &   A   \\
030328  &   1.52    &   140 &   47  $\pm$   3       &   328 $\pm$                   55      &   HETE每2 &   15  &   A   \\
030329  &   0.17    &   23  &   1.5 $\pm$   0.3     &   100 $\pm$                   23      &   HETE每2 &   15  &   A   \\
030429  &   2.65    &   10.3    &   2.16    $\pm$   0.26        &   128 $\pm$                   26      &   HETE每2 &   15  &   A   \\
030528  &   0.78    &   49.2    &   2.5 $\pm$   0.3     &   57  $\pm$                   9       &   HETE每2 &   10  &   A   \\
040912  &   1.563   &   143 &   1.3 $\pm$   0.3     &   44  $\pm$                   33      &   HETE每2 &   11  &   A   \\
040924  &   0.859   &   5   &   0.95    $\pm$   0.09        &   102 $\pm$                   35      &   HETE每2 &   15  &   A   \\
041006  &   0.716   &   25  &   3   $\pm$   0.9     &   98  $\pm$                   20      &   HETE每2 &   15  &   A   \\
050126A &   1.29    &   24.8    &   0.736   $\pm$   0.16        &   263 $\pm$                   110     &   Swift   &   16  &   A   \\
050223  &   0.5915  &   22.5    &   0.121   $\pm$   0.0177      &   110 $\pm$                   54      &   Swift   &   16  &   A   \\
050318  &   1.44    &   32  &   2.2 $\pm$   0.16        &   115 $\pm$                   25      &   Swift   &   15  &   A   \\
050401  &   2.9 &   33.3    &   35  $\pm$   7       &   467 $\pm$                   110     &   Swift   &   15  &   A   \\
050416A &   0.65    &   2.5 &   0.1 $\pm$   0.01        &   25.1    $\pm$                   4.2     &   Swift   &   15  &   A   \\
050505  &   4.27    &   58.9    &   17.6    $\pm$   2.61        &   661 $\pm$                   245     &   Swift   &   16  &   A   \\
050509B &   0.225   &   0.04    &   0.00024     $^{+0.00044}    _{-0.0001}$ &   102 $\pm$                   10      &   Swift   &   18  &   N   \\
050525A &   0.606   &   8.8 &   2.5 $\pm$   0.43        &   127 $\pm$                   10      &   Swift   &   15  &   A   \\
050603  &   2.821   &   12.4    &   60  $\pm$   4       &   1333    $\pm$                   107     &   Konus   &   15  &   A   \\
050709  &   0.16    &   0.07    &   0.0033  $\pm$   0.0001      &   97.4    $\pm$                   11.6        &   HETE-2  &   1   &   N   \\
050803  &   0.422   &   87.9    &   0.186   $\pm$   0.0399      &   138 $\pm$                   48      &   Swift   &   16  &   A   \\
050813  &   1.8 &   0.6 &   0.015       $^{+0.0025} _{-0.008}$  &   150 $\pm$                   15      &   Swift   &   18  &   N   \\
050814  &   5.3 &   150.9   &   11.2    $\pm$   2.43        &   339 $\pm$                   47      &   Swift   &   16  &   A   \\
050820  &   2.612   &   26  &   97.4    $\pm$   7.8     &   1325    $\pm$                   277     &   Konus   &   15  &   A   \\
050904  &   6.29    &   174.2   &   124 $\pm$   13      &   3178    $\pm$                   1094        &   Swift   &   15  &   A   \\
050908  &   3.344   &   19.4    &   1.97    $\pm$   0.321       &   195 $\pm$                   36      &   Swift   &   16  &   A   \\
050922C &   2.198   &   4.6 &   5.3 $\pm$   1.7     &   415 $\pm$                   111     &   HETE每2 &   15  &   A   \\
051022  &   0.8 &   200 &   54  $\pm$   5       &   754 $\pm$                   258     &   HETE每2 &   15  &   A   \\
051109A &   2.346   &   37.2    &   6.5 $\pm$   0.7     &   539 $\pm$                   200     &   Konus   &   15  &   A   \\
051221A &   0.547   &   1.4 &   0.3 $\pm$   0.04        &   621 $\pm$                   144     &   Swift   &   18  &   N   \\
060115  &   3.53    &   139.6   &   6.3 $\pm$   0.9     &   285 $\pm$                   34      &   Swift   &   15  &   A   \\
060124  &   2.297   &   750 &   41  $\pm$   6       &   784 $\pm$                   285     &   Konus   &   15  &   A   \\
\hline
\end{tabular}
\end{table*}

\setcounter{table}{0}
\begin{table*}
\caption{--- continued}
\begin{tabular}{llllllllllll}
\hline
GRB&z&$T_{90}$&$E_{iso}$&$E_{p,r}$&Instruments&Refs&Note\\
&&&$10^{52}$ erg&keV&&&\\
\hline
060206  &   4.048   &   7.6 &   4.3 $\pm$   0.9     &   394 $\pm$                   46      &   Swift   &   15  &   A   \\
060210  &   3.91    &   255 &   41.53   $\pm$   5.7     &   575 $\pm$                   186     &   Swift   &   16  &   A   \\
060218  &   0.0331  &   2100    &   0.0053  $\pm$   0.0003      &   4.9 $\pm$                   0.3     &   Swift   &   15  &   A   \\
060223A &   4.41    &   11.3    &   4.29    $\pm$   0.664       &   339 $\pm$                   63      &   Swift   &   16  &   A   \\
060418  &   1.489   &   103.1   &   13  $\pm$   3       &   572 $\pm$                   143     &   Konus   &   15  &   A   \\
060502B &   0.287   &   0.131   &   0.003       $^{+0.005}  _{-0.002}$  &   193 $\pm$                   19      &   Swift   &   18  &   N   \\
060510B &   4.9 &   275.2   &   36.7    $\pm$   2.87        &   575 $\pm$                   227     &   Swift   &   16  &   A   \\
060522  &   5.11    &   71.1    &   7.77    $\pm$   1.52        &   427 $\pm$                   79      &   Swift   &   16  &   A   \\
060526  &   3.21    &   298.2   &   2.6 $\pm$   0.3     &   105 $\pm$                   21      &   Swift   &   15  &   A   \\
060605  &   3.78    &   79.1    &   2.83    $\pm$   0.45        &   490 $\pm$                   251     &   Swift   &   16  &   A   \\
060607A &   3.082   &   102.2   &   10.9    $\pm$   1.55        &   575 $\pm$                   200     &   Swift   &   16  &   A   \\
060614  &   0.125   &   108.7   &   0.21    $\pm$   0.09        &   55  $\pm$                   45      &   Konus   &   15  &   A   \\
060707  &   3.43    &   66.2    &   5.4 $\pm$   1       &   279 $\pm$                   28      &   Swift   &   15  &   A   \\
060714  &   2.711   &   115 &   13.4    $\pm$   0.912       &   234 $\pm$                   109     &   Swift   &   16  &   A   \\
060814  &   0.84    &   145.3   &   7   $\pm$   0.7     &   473 $\pm$                   155     &   Konus   &   15  &   A   \\
060904B &   0.703   &   171.5   &   0.364   $\pm$   0.0743      &   135 $\pm$                   41      &   Swift   &   16  &   A   \\
060906  &   3.686   &   43.5    &   14.9    $\pm$   1.56        &   209 $\pm$                   43      &   Swift   &   16  &   A   \\
060908  &   2.43    &   19.3    &   9.8 $\pm$   0.9     &   514 $\pm$                   102     &   Swift   &   15  &   A   \\
060927  &   5.6 &   22.5    &   13.8    $\pm$   2       &   475 $\pm$                   47      &   Swift   &   15  &   A   \\
061006  &   0.4377  &   129.9   &   0.2 $\pm$   0.03        &   955 $\pm$                   267     &   Swift   &   1   &   N   \\
061007  &   1.261   &   75.3    &   86  $\pm$   9       &   890 $\pm$                   124     &   Konus   &   15  &   A   \\
061121  &   1.314   &   81.3    &   22.5    $\pm$   2.6     &   1289    $\pm$                   153     &   Konus   &   15  &   A   \\
061126  &   1.1588  &   70.8    &   30  $\pm$   3       &   1337    $\pm$                   410     &   Swift   &   17  &   A   \\
061201  &   0.111   &   0.76    &   3       $^{+4}  _{-2}$  &   969 $\pm$                   508     &   Swift   &   18  &   N   \\
061217  &   0.827   &   0.21    &   0.03        $^{+0.04}   _{-0.02}$   &   216 $\pm$                   22      &   Swift   &   18  &   N   \\
061222B &   3.355   &   40  &   10.3    $\pm$   1.6     &   200 $\pm$                   28      &   Swift   &   16  &   A   \\
070110  &   2.352   &   88.4    &   5.5 $\pm$   1.5     &   370 $\pm$                   170     &   Swift   &   8   &   A   \\
070125  &   1.547   &   70  &   80.2    $\pm$   8       &   934 $\pm$                   148     &   Konus   &   15  &   A   \\
070429B &   0.904   &   0.47    &   0.07        $^{+0.11}   _{-0.02}$   &   813 $\pm$                   81      &   Swift   &   18  &   N   \\
070508  &   0.82    &   21.2    &   8       $^{+2}  _{-1}$  &   378.56      $^{+    138.32  }   _{- 74.62   }$  &   Swift   &   14  &   A   \\
070714B &   0.92    &   2   &   1.1 $\pm$   0.1     &   2150    $\pm$                   1113        &   Swift   &   1   &   N   \\
070724A &   0.457   &   0.4 &   0.00245     $^{+0.00175}    _{-0.0055}$ &   119 $\pm$                   12      &   Swift   &   18  &   N   \\
070809  &   0.2187  &   1.3 &   0.00131     $^{+0.00103}    _{-0.00285}$    &   91  $\pm$                   9       &   Swift   &   18  &   N   \\
071003  &   1.604   &   150 &   36  $\pm$   4       &   2077    $\pm$                   286     &   Swift   &   19  &   A   \\
071010B &   0.947   &   35.7    &   1.7 $\pm$   0.9     &   101 $\pm$                   20      &   Konus   &   15  &   A   \\
071020  &   2.145   &   4.2 &   9.5 $\pm$   4.3     &   1013    $\pm$                   160     &   Konus   &   15  &   A   \\
071117  &   1.331   &   6.6 &   4.1 $\pm$   0.9     &   647 $\pm$                   226     &   Konus   &   12  &   A   \\
071227  &   0.383   &   1.8 &   0.1 $\pm$   0.02        &   1384    $\pm$                   277     &   Swift   &   3   &   N   \\
080319B &   0.937   &   50  &   114 $\pm$   9       &   1261    $\pm$                   65      &   Swift   &   15  &   A   \\
080319C &   1.95    &   34  &   14.1    $\pm$   2.8     &   906 $\pm$                   272     &   Swift   &   15  &   A   \\
080411  &   1.03    &   56  &   15.6    $\pm$   0.9     &   524 $\pm$                   70      &   Konus   &   13  &   A   \\
080413A &   2.433   &   46  &   8.1 $\pm$   2       &   584 $\pm$                   180     &   Swift   &   4   &   A   \\
080413B &   1.1 &   8   &   2.4 $\pm$   0.3     &   150 $\pm$                   30      &   Swift   &   5   &   A   \\
080514B &   1.8 &   7   &   17  $\pm$   4       &   627 $\pm$                   65      &   Konus   &   19  &   A   \\
080603B &   2.69    &   60  &   11  $\pm$   1       &   376 $\pm$                   100     &   Konus   &   19  &   A   \\
080605  &   1.6398  &   20  &   24  $\pm$   2       &   650 $\pm$                   55      &   Konus   &   19  &   A   \\
080607  &   3.036   &   79  &   188 $\pm$   10      &   1691    $\pm$                   226     &   Konus   &   19  &   A   \\
080721  &   2.591   &   16.2    &   126 $\pm$   22      &   1741    $\pm$                   227     &   Swift   &   19  &   A   \\
080804  &   2.2 &   34  &   11.5    $\pm$   2       &   810 $\pm$                   45      &   Swift   &   8   &   A   \\
080810  &   3.35    &   106 &   45  $\pm$   5       &   1470    $\pm$                   180     &   Swift   &   19  &   A   \\
080913  &   6.7 &   8   &   8.6 $\pm$   2.5     &   710 $\pm$                   350     &   Konus   &   19  &   A   \\
080916A &   0.689   &   60  &   2.27    $\pm$   0.76        &   184.101 $\pm$                   15.201      &   Swift   &   6   &   A   \\
080916B &   4.35    &   32  &   563             &   424 $\pm$                   24      &   Swift   &   7   &   A   \\
080916C &   4.35    &   62.977  &   387 $\pm$   46      &   2268.4  $\pm$                   128.4       &   Fermi   &   6   &   A   \\
081007  &   0.5295  &   10  &   0.16    $\pm$   0.03        &   61  $\pm$                   15      &   Swift   &   19  &   A   \\
081028  &   3.038   &   260 &   17  $\pm$   2       &   234 $\pm$                   93      &   Swift   &   19  &   A   \\
081118  &   2.58    &   67  &   4.3 $\pm$   0.9     &   147 $\pm$                   14      &   Swift   &   19  &   A   \\
081121  &   2.512   &   14  &   26  $\pm$   5       &   871 $\pm$                   123     &   Swift   &   19  &   A   \\
081203A &   2.1 &   294 &   35  $\pm$   3       &   1541    $\pm$                   757     &   Swift   &   8   &   A   \\
081222  &   2.77    &   24  &   30  $\pm$   3       &   505 $\pm$                   34      &   Swift   &   19  &   A   \\
090102  &   1.547   &   27  &   22  $\pm$   4       &   1149    $\pm$                   166     &   Konus   &   19  &   A   \\
090323  &   3.57    &   135.17  &   410 $\pm$   50      &   1901    $\pm$                   343     &   Konus   &   19  &   A   \\
\hline
\end{tabular}
\end{table*}

\setcounter{table}{0}
\begin{table*}
\caption{--- continued}
\begin{tabular}{llllllllllll}
\hline
GRB&z&$T_{90}$&$E_{iso}$&$E_{p,r}$&Instruments&Refs&Note\\
&&&$10^{52}$ erg&keV&&&\\
\hline
090328  &   0.736   &   61.697  &   13  $\pm$   3       &   1028    $\pm$                   312     &   Konus   &   19  &   A   \\
090418  &   1.608   &   56  &   16  $\pm$   4       &   1567    $\pm$                   384     &   Swift   &   19  &   A   \\
090423  &   8.1 &   10.3    &   11  $\pm$   3       &   491 $\pm$                   200     &   Swift   &   19  &   A   \\
090424  &   0.544   &   48  &   4.6 $\pm$   0.9     &   273 $\pm$                   50      &   Swift   &   19  &   A   \\
090425A &   0.544   &   75.393  &   4.48                &   177 $\pm$                   3       &   Fermi   &   7   &   A   \\
090426  &   2.609   &   1.2 &   0.42        $^{+0.38}   _{-1.14}$   &   177 $\pm$                   82      &   Swift   &   18  &   A   \\
090510  &   0.903   &   0.3 &   3.75    $\pm$   0.25        &   8370    $\pm$                   760     &   Swift   &   6   &   N   \\
090516  &   4.109   &   210 &   88.5    $\pm$   19.2        &   948.2304        $^{+    502.73  }   _{- 217.1325    }$  &   Swift   &   6   &   A   \\
090618  &   0.54    &   113.2   &   25.4    $\pm$   0.6     &   239.47      $^{+        }   _{- 16.94   }$  &   Swift   &   6   &   A   \\
090812  &   2.452   &   66.7    &   40.3    $\pm$   4       &   2023    $\pm$                   663     &   Swift   &   8   &   A   \\
090902B &   1.822   &   19.328  &   305 $\pm$   2       &   2187.05 $\pm$                   31.042      &   Fermi   &   6   &   A   \\
090926A &   2.1062  &   13.76   &   186 $\pm$   5       &   975.3468    $\pm$                   12.4248     &   Fermi   &   6   &   A   \\
090926B &   1.24    &   109.7   &   3.55    $\pm$   0.12        &   203.84  $\pm$                   4.48        &   Swift   &   6   &   A   \\
091003A &   0.8969  &   20.224  &   10.6    $\pm$   0.1     &   922.2728    $\pm$                   44.76684        &   Fermi   &   6   &   A   \\
091020  &   1.71    &   34.6    &   12.2    $\pm$   2.4     &   129.809 $\pm$                   19.241      &   Swift   &   6   &   A   \\
091024  &   1.092   &   109.8   &   28  $\pm$   3       &   794 $\pm$                   231     &   Swift   &   8   &   A   \\
091029  &   2.752   &   39.2    &   7.4 $\pm$   0.74        &   230 $\pm$                   66      &   Swift   &   8   &   A   \\
091127  &   0.49    &   7.1 &   1.63    $\pm$   0.02        &   53.64   $\pm$                   2.98        &   Swift   &   6   &   A   \\
091208B &   1.063   &   14.9    &   2.01    $\pm$   0.07        &   297.4846        $^{+    37.13   }   _{- 28.6757 }$  &   Swift   &   6   &   A   \\
100117A &   0.92    &   0.3 &   0.09    $\pm$   0.01        &   551     $^{+    142.00  }   _{- 96  }$  &   Swift   &   6   &   N   \\
100414A &   1.368   &   26.497  &   76.6    $\pm$   1.2     &   1486.157        $^{+    29.60   }   _{- 28.6528 }$  &   Fermi   &   6   &   A   \\
100621A &   0.542   &   63.6    &   4.37    $\pm$   0.5     &   146 $\pm$                   23.1        &   Swift   &   8   &   A   \\
100728B &   2.106   &   12.1    &   2.66    $\pm$   0.11        &   406.886 $\pm$                   46.59       &   Swift   &   6   &   A   \\
100814A &   1.44    &   174.5   &   14.8    $\pm$   0.5     &   259.616     $^{+    33.92   }   _{- 30.744  }$  &   Swift   &   6   &   A   \\
100816A &   0.8049  &   2.9 &   0.73    $\pm$   0.02        &   246.7298    $\pm$                   8.48303     &   Swift   &   6   &   A   \\
100906A &   1.727   &   114.4   &   28.9    $\pm$   0.3     &   289.062     $^{+    47.72   }   _{- 55.0854 }$  &   Swift   &   6   &   A   \\
101219A &   0.718   &   0.6 &   0.49    $\pm$   0.07        &   842     $^{+    177.00  }   _{- 136 }$  &   Swift   &   2   &   N   \\
101219B &   0.55    &   34  &   0.59    $\pm$   0.04        &   108.5   $\pm$                   12.4        &   Swift   &   6   &   A   \\
110205A &   2.22    &   257 &   56  $\pm$   6       &   715 $\pm$                   239     &   Swift   &   8   &   A   \\
110213A &   1.46    &   48  &   6.9 $\pm$   0.2     &   242.064     $^{+    20.91   }   _{- 16.974  }$  &   Swift   &   6   &   A   \\
\hline
\end{tabular}\label{tab1}
\begin{flushleft}
\textbf{Note.} The ``Note" column: `A' represents Amati type bursts,
and `N' represents non-Amati type bursts. References: (1)Ghirlanda
et al. 2008 and therein; (2) Golenetskii et al. 2010; (3)
Golenetskii et al. (2007a); (4) Ohno et al. (2008); (5) Barthelmy et
al. (2008); (6) Zhang et al. (2012) and therein; (7) Ghirlanda et
al. (2010) and therein; (8) Ghirlanda et al. (2012) and therein; (9)
Ulanov et al. (2005); (10) Sakamoto et al. (2005) and therein; (11)
Stratta et al. (2007); (12) Golenetskii et al. (2007b); (13)
Golenetskii et al. (2008); (14) Nava et al. (2008); (15) Amati et
al. (2008) and therein; (16) Cabrera et al. (2007) and therein; (17)
Perley et al. (2008); (18) Butler et al. (2007) and therein; (19)
Amati et al. (2009) and therein.
\end{flushleft}
\end{table*}


\begin{thebibliography}{99}
\bibitem[\protect\citeauthoryear{}{}]{}Amati L., Frontera F., Tavani M. et al., 2002, A\&A, 390, 81
\bibitem[\protect\citeauthoryear{}{}]{}Amati L., 2006, MNRAS, 372, 233
\bibitem[\protect\citeauthoryear{}{}]{}Amati L., Della Valle M., Frontera F., et al., 2007, A\&A, 463, 913
\bibitem[\protect\citeauthoryear{}{}]{}Amati L., Guidorzi C., Frontera F. et al., 2008, MNRAS, 391, 577
\bibitem[\protect\citeauthoryear{}{}]{}Amati L., Frontera F., Guidorzi C., 2009, A\&A, 508, 173
\bibitem[\protect\citeauthoryear{}{}]{}Amati L., 2010, Journal of the Korean Physical Society, 56, 1603
\bibitem[\protect\citeauthoryear{}{}]{}Barthelmy S. D., Chincarini G., Burrows D. N. et al., 2005, Nature, 438, 994
\bibitem[\protect\citeauthoryear{}{}]{}Berger E., Price P. A., Cenko S. B. et al., 2005, Nature, 438, 988
\bibitem[\protect\citeauthoryear{}{}]{}Barthelmy S. D., Chincarini G., Burrows D. N. et al., 2005, Nature, 438, 994
\bibitem[\protect\citeauthoryear{}{}]{}Band, D. L., \& Preece, R. 2005, ApJ, 627, 319
\bibitem[\protect\citeauthoryear{}{}]{}Butler N. R., Kocevski D., Bloom J. S., Curtis J. L., 2007, ApJ, 671, 656
\bibitem[\protect\citeauthoryear{}{}]{}Butler, N. R., Kocevski, D., \& Bloom, J. S. 2009, ApJ, 694, 76
\bibitem[\protect\citeauthoryear{}{}]{}Butler, N. R., Bloom, J. S., Poznanski, D. 2010, ApJ, 711, 495
\bibitem[\protect\citeauthoryear{}{}]{}Bosnjak, Z., Celotti, A., Longo, F., Barbiellini, G. 2008, MNRAS,384, 599
\bibitem[\protect\citeauthoryear{}{}]{}Barthelmy S. D., Baumgartner W., Cummings J. 2008, GCN 7606
\bibitem[\protect\citeauthoryear{}{}]{}Barthelmy S. D., Baumgartner W. H., Cummings J. R. et al. 2008, GCN, 8428
\bibitem[\protect\citeauthoryear{}{}]{}Cenko S. B., Kasliwal M., Harrison F. A., 2006, ApJ, 652, 490
\bibitem[\protect\citeauthoryear{}{}]{}Cabrera J. I., Firmani C., Avila-Reese V. et al., 2007, MNRAS, 382, 342
\bibitem[\protect\citeauthoryear{}{}]{}Della Valle M., Chincarini G., Panagia N. et al., 2006, Nature, 444, 1050
\bibitem[\protect\citeauthoryear{}{}]{}Eichler D., Livio M., Piran T. et al., 1989, Nature, 340, 126
\bibitem[\protect\citeauthoryear{}{}]{}Fishman G. J., Meegan C. A., 1995, ARA\&A, 33, 415
\bibitem[\protect\citeauthoryear{}{}]{}Fruchter A. S., Levan A. J., Strolger L. et al., 2006, Nature, 441, 463
\bibitem[\protect\citeauthoryear{}{}]{}Fynbo J. P. U., Watson D., Thone C. C. et al., 2006, Nature, 444, 1047
\bibitem[\protect\citeauthoryear{}{}]{}Ghirlanda, G., Ghisellini, G., Firmani, C., Celotti, A., \& Bosnjak,Z. 2005, MNRAS, 360, 45
\bibitem[\protect\citeauthoryear{}{}]{}Ghirlanda G., Nava L., Ghisellini G. et al., 2008, MNRAS, 387, 319
\bibitem[\protect\citeauthoryear{}{}]{}Ghirlanda G., Nava L., Ghisellini G. et al., 2009, A\&A, 496, 585
\bibitem[\protect\citeauthoryear{}{}]{}Ghirlanda G., Nava L., Ghisellini G. 2010, A\&A, 511, A43
\bibitem[\protect\citeauthoryear{}{}]{}Ghirlanda G., Nava L., Ghisellini G. et al., 2012, MNRAS, 420, 483
\bibitem[\protect\citeauthoryear{}{}]{}Gehrels N., Chincarini G., Giommi P. et al., 2004, \apj, 611, 1005
\bibitem[\protect\citeauthoryear{}{}]{}Gehrels N., Norris J. P., Barthelmy S. D. et al., 2006, Nature, 444, 1044
\bibitem[\protect\citeauthoryear{}{}]{}Gal-Yam A., Fox D. B., Price P. A. et al., 2006, Nature, 444, 1053
\bibitem[\protect\citeauthoryear{}{}]{}Goldstein A., Preece R. D., Briggs M. S., 2010, ApJ, 721, 1329
\bibitem[\protect\citeauthoryear{}{}]{}Golenetskii S., Aptekar R., Mazets E. et al., 2007a, GRB Circ., 7155, 1
\bibitem[\protect\citeauthoryear{}{}]{}Golenetskii S., Aptekar R., Mazets E. et al., 2007b, GCN 7114
\bibitem[\protect\citeauthoryear{}{}]{}Golenetskii S., Aptekar R., Mazets E. et al., 2008, GCN 7589
\bibitem[\protect\citeauthoryear{}{}]{}Golenetskii S., Aptekar R., Frederiks D., et al. 2010, GRB Circ., 11470, 1
\bibitem[\protect\citeauthoryear{}{}]{}Gruber D., Greiner J., von Kienlin A., et al. 2011, A\&A, 531, A20
\bibitem[\protect\citeauthoryear{}{}]{}Horvath I., 1998, \apj, 508, 757
\bibitem[\protect\citeauthoryear{}{}]{}Hjorth J., Sollerman J., Moller P. et al., 2003, Nature, 423, 847
\bibitem[\protect\citeauthoryear{}{}]{}Hjorth J., Watson D., Fynbo J. P. U. et al., 2005, Nature, 437, 859
\bibitem[\protect\citeauthoryear{}{}]{}Kouveliotou C., Meegan C. A., Fishman G. J., et al., 1993, \apj, 413, L101
\bibitem[\protect\citeauthoryear{}{}]{}Krimm, H. A., Yamaoka, K., Sugita, S., et al. 2009, ApJ, 704, 1405
\bibitem[\protect\citeauthoryear{}{}]{}Lloyd N. M., Petrosian V., Mallozzi R. S., 2000, ApJ, 534, 227
\bibitem[\protect\citeauthoryear{}{}]{}Lu H. J., Liang E. W., Zhang B.-B., Zhang, B., 2010, ApJ, 725, 1965
\bibitem[\protect\citeauthoryear{}{}]{}MacFadyen A. I., Woosley S. E., 1999, \apj, 524, 262
\bibitem[\protect\citeauthoryear{}{}]{}McBreen B., Hurley K. J., Long R., et al., 1994, MNRAS, 271, 662
\bibitem[\protect\citeauthoryear{}{}]{}Norris J. P., Bonnell J. T., 2006, ApJ, 643, 266
\bibitem[\protect\citeauthoryear{}{}]{}Nakar, E., \& Piran, T. 2005, MNRAS, 360, L73
\bibitem[\protect\citeauthoryear{}{}]{}Nava L., Ghirlanda G., Ghisellini G., Firmani C., 2008, MNRAS, 391, 639
\bibitem[\protect\citeauthoryear{}{}]{}Ohno M., Kokubun M., Suzuki M., Takahashi T. et al. 2008, GCN, 7630
\bibitem[\protect\citeauthoryear{}{}]{}Paczynski B., 1986, \apj, 308, L43
\bibitem[\protect\citeauthoryear{}{}]{}Paczynski B., 1998, \apj, 494, L45
\bibitem[\protect\citeauthoryear{}{}]{}Perley D. A. et al., 2008, ApJ, 672, 449
\bibitem[\protect\citeauthoryear{}{}]{}Piranomonte S., D＊Avanzo P., Covino S., et al. 2008, A\&A, 491, 183
\bibitem[\protect\citeauthoryear{}{}]{}Qin Y.-P., Xie G.-.Z, Xue S.-J. et al., 2000, PASJ, 52, 759
\bibitem[\protect\citeauthoryear{}{}]{}Qin Y.-P., Gupta A. C., Fan J.-H., Su C.-Y., Lu R.-J., 2010, Science China G, 53, 1375
\bibitem[\protect\citeauthoryear{}{}]{}Stanek K. Z., Matheson T., Garnavich P. M. et al., 2003, \apj, 591, L17
\bibitem[\protect\citeauthoryear{}{}]{}Stratta G. et al., 2007, A\&A, 461, 485
\bibitem[\protect\citeauthoryear{}{}]{}Ulanov M. V., Golenetskii S. V., Frederiks D. D. et al., 2005, Nuovo Cimento C, 28, 351
\bibitem[\protect\citeauthoryear{}{}]{}Woosley S. E., 1993, \apj, 405, 273
\bibitem[\protect\citeauthoryear{}{}]{}Zhang B., Meszaros P., 2002, ApJ, 581, 1236
\bibitem[\protect\citeauthoryear{}{}]{}Zhang B., Zhang B.-B., Liang E.-W. et al., 2007, ApJ, 655, L25
\bibitem[\protect\citeauthoryear{}{}]{}Zhang F.-W., Shao L., Yan J.-Z., Wei D.-M., 2012, ApJ, 750, 88

\end{thebibliography}
\end{document}